**Title:** *Adaptive antenna system by ESP32-PICO-D4 and its application to web radio system*

**Authors:** Toshiro Kodera

**Affiliations:** Meisei University

**Contact email:** *toshiro.kodera@meisei-u.ac.jp*

**Abstract:** *Adaptive antenna technique has an important role in the IoT environment in order to establish reliable and stable wireless communication in high congestion situation. Even if knowing antenna characteristics in advance, electromagnetic wave propagation in non-line-of-sight environment is very complex and unpredictable, therefore, the adjustment the antenna radiation for the optimum signal reception is important for the better wireless link. This article presents a simple but effective adaptive antenna system for Wi-Fi utilizing the function of a highly integrated component, ESP32-PICO-D4. This chip is a system-in-chip containing all components for Wi-Fi and Bluetooth application except for antenna. Together with SP3T RF switch and dielectric antennas and high-resolution audio DAC, completed web-radio system is made in the size of 50 x 50 mm.*

**Keywords:** *beam switching, adaptive antenna, system-in-chip, ESP32, web-radio*

**Specifications table**

| Hardware name | *Web radio with adaptive antenna* |
|---|---|
| Subject area | • Engineering and Material Science |
| Hardware type | • Electrical engineering and computer science |
| Open Source License | *Creative Commons Attribution-Share Alike license* |
| Cost of Hardware | *30USD* |
| Source File Repository | *https://osf.io/rf8d7/* |

# 1. Hardware in context

1.1 Introduction

The recent drastic increase of IoT edge device leads to high congestion in the wireless communication. One solution for this issue is an adaptive beamforming. Adaptive beamforming and adaptive antenna techniques already have a long history and its effectiveness are well confirmed, nevertheless, they are rarely adapted for consumer product due to cost and circuit complexity.

Recently, the significant progress of semiconductor integrated circuit technology has enabled RF transceiver for Wi-Fi and Bluetooth in one package and eventually very powerful dual-core processor, flash memory for storage, SRAM for main memory and all peripherals are integrated into single package. The main chip of this article, ESP32-PICO-D4 is one kind of such a highly integrated chip. By using this chip, not only RF transceiver implementation but also fully programmable system is easily realized.

This article presents a simple but effective adaptive antenna system for Wi-Fi utilizing programmability of the chip. Together with SP3T RF switch and dielectric antennas and high-resolution audio DAC, completed web-radio system is made in the size of 50 x 50 mm. As demonstrated in this article, the circuit will select one antenna for maximum RSSI to a given access point in the initial boot process and move to start the web radio system. The web radio system is based on two sophisticated codes [9][10] and station preset and selection can be made through a web browser.

The application of the implementation of adaptive antenna selection introduced in this article is not limited a web radio system but also various IoT application is expected.

1.2 Comparison to previously published designs

ESP32 series have a lot of users all over the world but the author could not find this type of application, combining with RF SP3T switch in the literature. The adaptive antenna system is widely reported in various articles but the two-chip configuration consisting of the system-on-chip and SP3T switch is the simplest implementation.

1.3 Comparison to commercial instruments

The adaptive antenna is not novel technology but rarely applied to consumer product from the viewpoint of cost. ESP32-PICO-D4 is a very recent chip (released on Aug. 2016) and at the moment of submission, just one simple prototyping board (Espressif's genuine board, ESP32-PICO-KIT V3) can be found on the market. The circuit in this article costs just $30 even including audio DAC, which is quite competitive to established commercial products.

2. **Hardware description.**

*2.1 System overview*

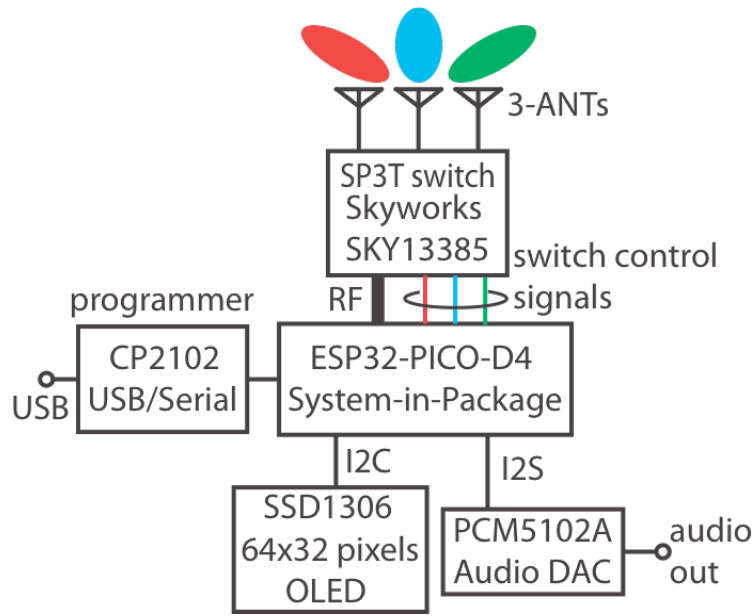

Fig.1 Block diagram of the whole system

Figure 1 shows the whole system block diagram. The main processor is ESP32-PICO-D4, which is a system-in-package component. ESP32-PICO-D4 contains all components required for Wi-Fi and Bluetooth application except for antenna just in 7 x 7 mm package. The programming environment is well organized by Espressif, the manufacturer of ESP32-PICO-D4 and released on GitHub [1]. Table 1 denotes the summary of ESP32-PICO-D4 specifications. Together with RF part, the high-performance main processor is able to support complex applications.

Table 1 Brief summary of ESP32-PICO-D4 speficiation[2]

| Items | Specification |
|---|---|
| Main Processor | 32bit LX6 Dual core 240 MHz |
| Containing RF radio | WiFi Bluetooth |
| SRAM | 520 kB |
| Flash Memory | 4 MB |
| Package size | 7 x 7 mm |

As shown in Fig. 1, the system has four peripherals including CP2102 USB serial interface [3], SSD1306 OLED display [4], PCM5102A high-resolution audio DAC [5] and antenna part. CP2102 works as a programmer of ESP32-PICO-D4 and also serial terminal. SSD1306 OLED display is connected through the I2C bus of ESP32-PICO-D4. OLED display works as data display of the internal data of ESP32-PICO-D4.

PCM5102A audio DAC is connected to ESP32-PICO-D4 via I2S bus and it can support up to 32-bit audio data of 384kHz sampling rate with 112 dB SNR [5]. In the system, PCM5102A works for MP3 streaming audio data decoder controlled by ESP32-PICO-D4. The antenna part is cored by Skyworks SKY13385-460LF SP3T switch, which supports RF signal up to 3.5 GHz [6]. Three outputs are connected to dielectric antennas through impedance matching circuit. The control signal terminal of SKY13385 is connected to GPIO of ESP32-PICO-D4 for software antenna radiation beam switching. More detailed design information and evaluation results are presented in the next section.

2.2 Beam switchable antenna system

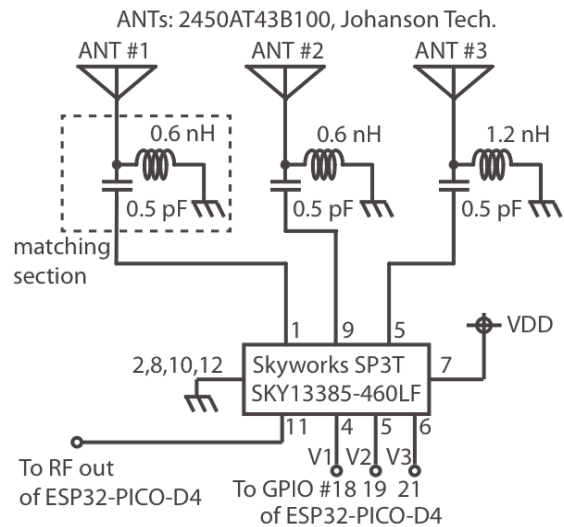

Fig. 2 Antenna switching part schematic including three antennas and SP3T switch.

Before embedding antenna part to the whole system, the sole antenna part consisting of an SP3T switch and dielectric antenna is designed and evaluated. Fig. 2 shows the antenna switching part schematic including three antennas and SP3T switch. Three dielectric antennas are 2450AT43B100 by Johanson technology [7]. Fig. 3 illustrates the solo antenna part PCB layout for characteristics evaluation.

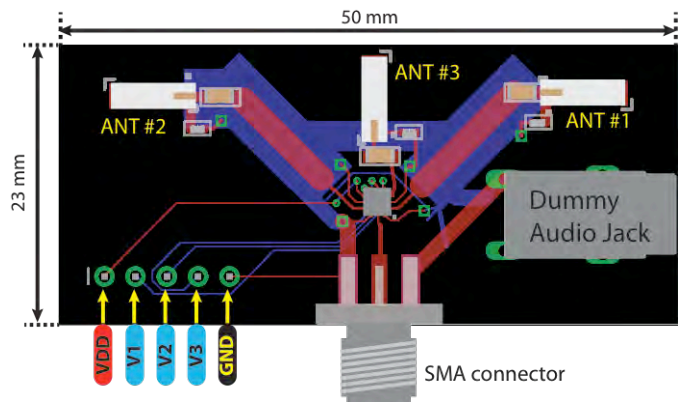

Fig. 3 PCB layout of antenna testing board corresponding to the schematic in Fig. 2.

As shown in Fig. 2, the PCB contains "Dummy Audio Jack" which is not a member of antenna part. The distance between dielectric antenna element (ANT #1) and the audio jack is very close enough to have an impact on the radiation characteristics, therefore, the performance of antenna part is optimized and evaluated with this object.

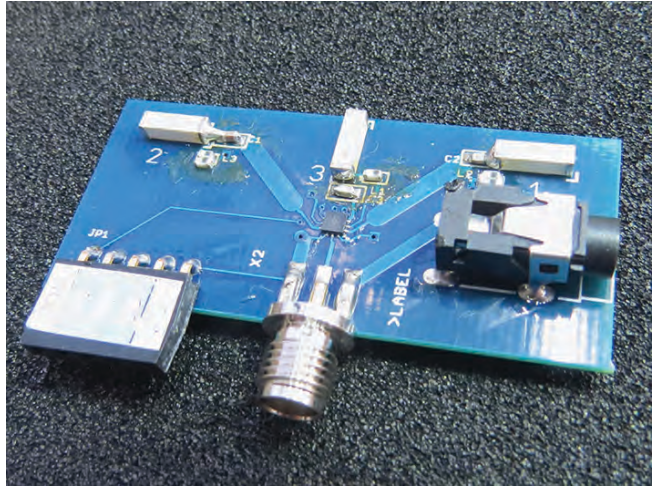

Fig. 4 Picture of antenna testing board.

Figure 4 shows the antenna testing board. In general, the impedance matching circuit should be designed according to each PCB layout [8]. As written in Fig. 2, inductance and capacitance are inserted at the terminal of an antenna element. These inductance and capacitance parameters are determined by measuring reflection coefficient using vector network analyzer for lowest reflection on 2.4 GHz ISM band.

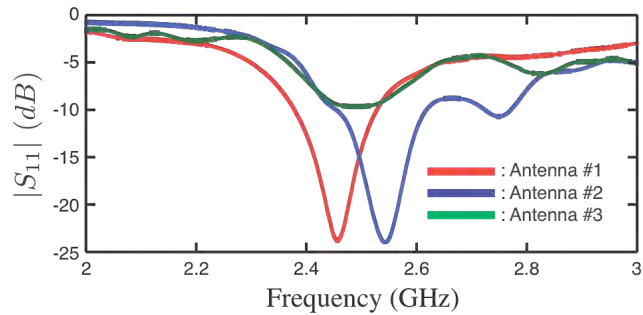

Fig. 5 Measured reflection coefficient of the antenna prototype in Fig. 4 for different antenna selection by SP3T switch.

Figure 5 shows the measured reflection coefficient of the antenna prototype in fig. 4 for different antenna selection by the SP3T switch. As shown in fig. 5, this dielectric antenna element has relatively narrow bandwidth with matching section. As it can be seen, antenna performance of antenna 1 and 2 has frequency shift even with the symmetric arrangement as is in Fig. 4. This shift can be attributed to the existence of near-by "dummy audio jack" but this characteristic shift can be utilized for wider frequency operation co-operated by antenna #1 (lower band) and #2 (higher band). The antenna #3 could not be fully optimized for larger return loss but it exhibits -10 dB around 2.4 GHz ISM band.

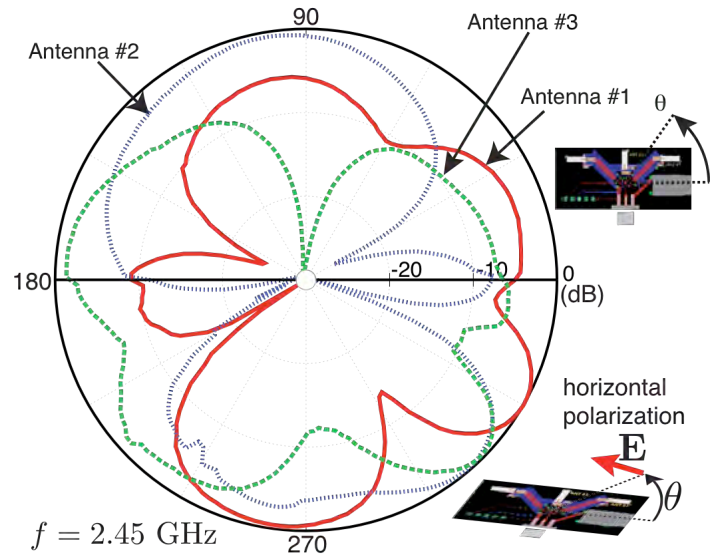

Fig. 6 Measured radiation pattern by each antenna element.

Figure 6 shows the measured radiation pattern by each antenna element. These patterns are measured for the horizontal polarization with changing $\theta$ as an explained inset. As it can be predicted by its structural symmetry, deep null can be observed around 180 deg for antenna #1 and 0 deg for antenna #2. In each case, the other antenna can complementary cover null radiation. The antenna #3 also covers the other antennas null radiation so very efficient adaptive radiation control can be expected by switching these three antennas depending on the communication condition. This antenna part is integrated into the whole system described in the next session.

*2.3 Integration to web-radio system*

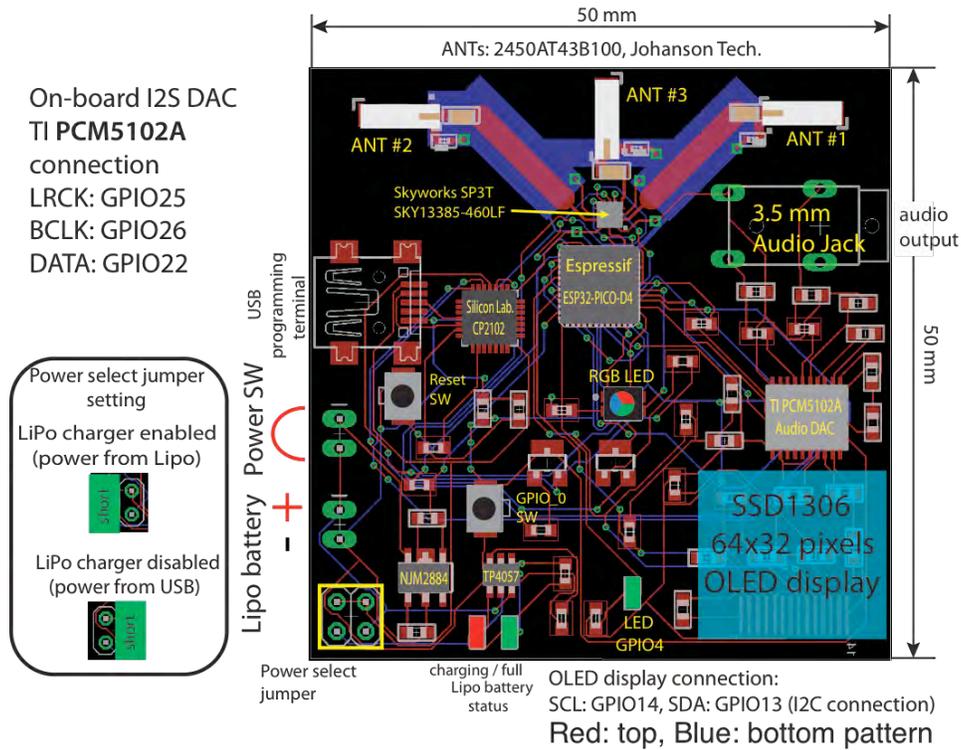

Fig. 7 PCB layout of whole system including antenna part, ESP32-PICO-D4, I2S DAC, OLED display, USB-Serial interface and power supply circuits.

Figure 7 shows the whole system PCB layout with 50 x 50 mm size. The antenna part in the previous section is integrated with RF I/O of ESP32-PICO-D4 and three switch control lines of SP3T is connected with GPIO #18, #19, #21 of ESP32-PICO-D4. This control line is also connected with RGB LED in order to indicate antenna selection state by three colors. As the power supply, this board support USB bus power or Lipo battery operation and this operation can be selected by onboard jumper, as explained inset. PCM5102A DAC is connected via I2S but through LRCK: GPIO 25, BCLK: GPIO26, and DATA: GPIO 22 of ESP32-PICO-D4. The audio output of PCM5102A is connected to 3.5 mm stereo audio jack. SSD1306 OLED display is connected to ESP32-PICO-D4 via I2C but through SCL: GPIO 14 and SDA: GPIO 13. ESP32 series has very flexibility for I2S and I2C bus pin selection and their arrangement can be programmed by software.

### 2.3.1 Caution for Lipo battery operation
The energy density of Lipo battery far exceeds conventional NiCd or NiMH battery, therefore, short circuit or reverse connection is extremely dangerous. In addition, wrong selection of jumper on this board may lead to the dangerous situation. When on board Lipo charger disabled, never connect Lipo battery to the onboard battery connector. Lipo battery connection with "charger disabled" setting will apply 5 V to Lipo battery, which is beyond the maximum rating of the battery and it is quite dangerous.

*2.4 ESP32-PICO-D4 firmware*

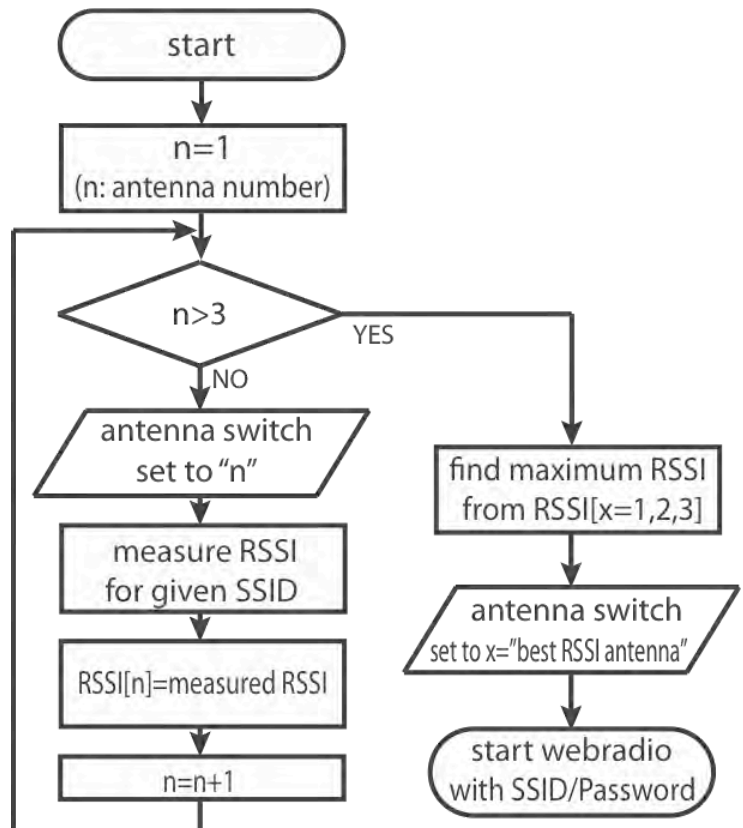

Fig. 8 Flow chart of the initial boot process of the system, where the optimum antenna element is selected by RSSI value for given SSID.

The completed system of this article realizes "web-radio" but before starting the web-radio application, antenna selection process is performed as shown in Fig. 8. As far as the author knows, RSSI of the established connection is not available from current Wi-Fi API in esp-idf [1], but during Wi-Fi scanning, RSSI of each SSID can be obtained from API. Therefore, the actual operation of "measure RSSI for given SSID" block of left side in Fig. 8 is performed by the following source code ( line 112 of app_main.c in "main" directory)

```c
ssid2=(const    char*)list[i].ssid;
if(strcmp(ssid2,ssid)==0){
    printf("rssi: %d\n", list[i].rssi);
    ant_rssi[antenna_num]=list[i].rssi;
}
```

, where "ssid2" is the SSID obtained during scanning and "ssid" is predefined and to be connected access point's SSID. Internally RSSI of all access points are obtained but only the RSSI in need is stored in ant_rssi[n]. Sweeping number "n" by the number of the antenna (here it is 3), RSSI for each antenna selection is obtained and the best antenna for connection link is determined.

After best antenna selection, web-radio program running on ESP32-PICO-D4 starts. The original source code implementing web-radio on ESP32 was written by MrBuddyCasino[9] and station presetting interface through web browser was implemented by n24bass[10]. The author of this article made an addition of OLED display support and the antenna switching and finding best antenna for communication link. The full source code is linked in this article.

3. Design files

**Design Files Summary**

| Design file name | File type | Open source license | Location of the file |
|---|---|---|---|
| *ESP32_BEAM.pdf* | *Circuit schematic in pdf format* | *CC-BY-SA 4.0* | *htttps://osf.io/rf8d7/* |
| *ESP32_BEAM.sch* | Circuit schematic in Eagle file format | *CC-BY-SA 4.0* | *htttps://osf.io/rf8d7/* |
| ESP32_BEAM.brd | PCB layout in Eagle file format | *CC-BY-SA 4.0* | *htttps://osf.io/rf8d7/* |
| ESP32_BEAM.zip | Firmware source codes for esp-idf | *CC-BY-SA 4.0* | *htttps://osf.io/rf8d7/* |

The following design files are provided with this article:

- ESP32_BEAM.pdf: Complete circuit schematic (pdf exported from Eagle cad)

- ESP32_BEAM.sch: Complete circuit schematic file in Eagle file format

- ESP32_BEAM.brd: Complete PCB layout file in Eagle file format

- ESP32_BEAM.zip: This zip file contains all of required source code for the system in this article. Firmware should be complied in Esprssif's programming environment, esp-idf, which can be found in [1].

## 4. Bill of Materials

| Designator | Component | Number | Cost per unit ($) | Total cost ($) | Source of materials | Material type |
|---|---|---|---|---|---|---|
| R2, R5-R6, R15 | Resistor, 1.5k, 0603 | 4 | 0.11 | 0.44 | Digikey RR08P1.5KDCT-ND | Composite |
| R8-R9 | Resistor, 470, 0603 | 2 | 0.11 | 0.22 | Digikey RR08P470DCT-ND | Composite |
| R11 | Resistor, 220k, 0603 | 1 | 0.11 | 0.11 | Digikey RR08P220KDCT-ND | Composite |
| R1, R3-R4, R7, R10, R12-R14 | Resistor, 10k, 0603 | 8 | 0.11 | 0.88 | Digikey RR08P10.0KDCT-ND | Composite |
| C8-C9 | Capacitor, 2.2uF, 16V, 0603 | 2 | 0.13 | 0.26 | Digikey 490-3296-1-ND | Ceramic |
| C16-C17 | Capacitor, 2200pF, 50V, 0603 | 2 | 0.16 | 0.32 | Digikey 490-4931-1-ND | *Ceramic* |
| C1-C7, C10-C15, C18-C19 | Capacitor, 1uF, 16V, 0603 | 15 | 0.1 | 1.5 | Digikey 587-1251-1-ND | Ceramic |
| C20-C21 | Capacitor, 0.5pF, 100V, 0603 | 2 | 0.57 | 1.14 | Digikey 490-3551-1-ND | Ceramic |
| L1-L3 | Inductor, 0.6nH | 2 | 0.1 | 0.2 | Digikey 490-6707-ND | Composite |
| LED4 | RGB LED, ASMB-MTB0-0A3A2 | 1 | 0.63 | 0.63 | Digikey 516-3279-1-ND | semi-conductor |
| LED1, LED3 | GREEN LED, LTST-C190KGKT | 2 | 0.27 | 0.54 | Digikey 160-1435-1-ND | semi-conductor |
| LED2 | RED LED, LTST-C191KRKT | 1 | 0.29 | 0.29 | Digikey 160-1447-1-ND | semi-conductor |
| U3 | ESP32-PICO-D4 | 1 | 4.5 | 4.5 | Electrodragon | semi-conductor |
| U2 | TI PCM5102A | 1 | 5.73 | 5.73 | Digikey 296-36707-1-ND | semi-conductor |
| Q1-Q2 | NPN, ZXTN04120HFFTA | 2 | 0.65 | 1.3 | Digikey ZXTN04120HFFCT-ND | semi-conductor |
| IC1 | CP2102N-A01-GQFN28 | 1 | 1.4 | 1.4 | Digikey 336-3694-ND | semi-conductor |
| U1 | TP4057-42-SOT26-R | 1 | 0.1 | 0.1 | Alibaba.com | semi-conductor |
| U6 | NJM2884U1-33-TE1 | 1 | 0.69 | 0.69 | Allied Electronics R1030663 | semi-conductor |
| LCD1 | SSD1306 64x32 OLED display | 1 | 3.66 | 3.66 | Aliexpress.com | Composite |
| ANT1-ANT3 | Antenna, 2450AT43B100 | 3 | 0.78 | 2.34 | Digikey 712-1010-1-ND | Ceramic |
| X1 | 3.5mm stereo jack 1503_08 | 1 | 1.07 | 1.07 | Allied Electronics 70151460 | Composite |
| CN1 | mini USB type B SMD | 1 | 0.54 | 0.54 | Digikey 151-1206-1-ND | Composite |
| S1-S2 | tactile SW, PTS820 J15M | 2 | 0.44 | 0.88 | Digikey CKN10506CT-ND | Composite |

**The total components cost $28.84. PCB can be ordered to prototyping PCB service company less than $5 per board. Total cost of one board is less than $35.**

## 5. Build Instructions

*5.1 PCB fabrication*

By using Autodesk Eagle, the files required by PCB prototyping service company can be generated from the attached Eagle files (ESP32_BEAM.brd and ESP32_BEAM.sch). The minimum separation is 0.5 mm pitch of ESP32-PICO-D4's QFN48 package, which has well margin to all of PCB service.

5.2 Soldering process

General stencil-reflow process is enough to complete PCB.

5.3 Firmware compiling and uploading

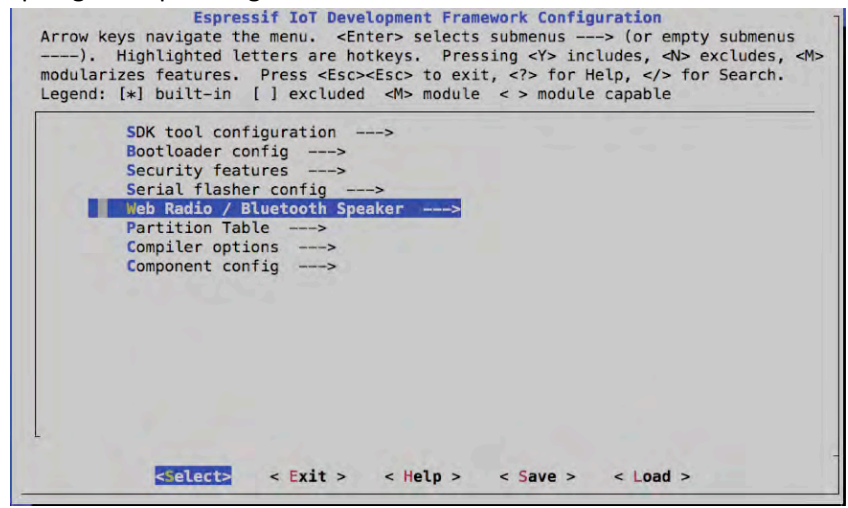

Fig. 9 esp-idf configuration window

Executing "make menuconfig" under correctly installed esp-idf environment will provide the configuration window as shown in Fig. 9. In the submenu of "Web Radio / Bluetooth Speaker", SSID and password of the access point can be configured. After writing SSID/password, simply exit configuration menu and executing "make; make flash" will flash the firmware including all functions (Antenna selection, OLED, web interface, web-radio).

6. Operation Instructions

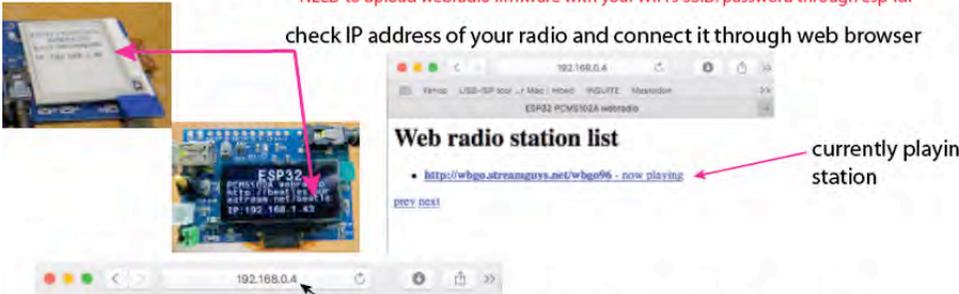

Fig. 10 Station pre-setting through web browser

After the boot process of the board including best antenna selection, web-radio will start. This web interface extension is implemented by n24bass[10] and fundamental part of web-radio is implemented by MrBuddyCasino[9]. As fully explained in Fig. 10, presetting station and its selection can be done through web interface. Not only the web interface but also pushing on-board GPIO 0 switch will change the preset station sequentially.

**7. Validation and Characterization**

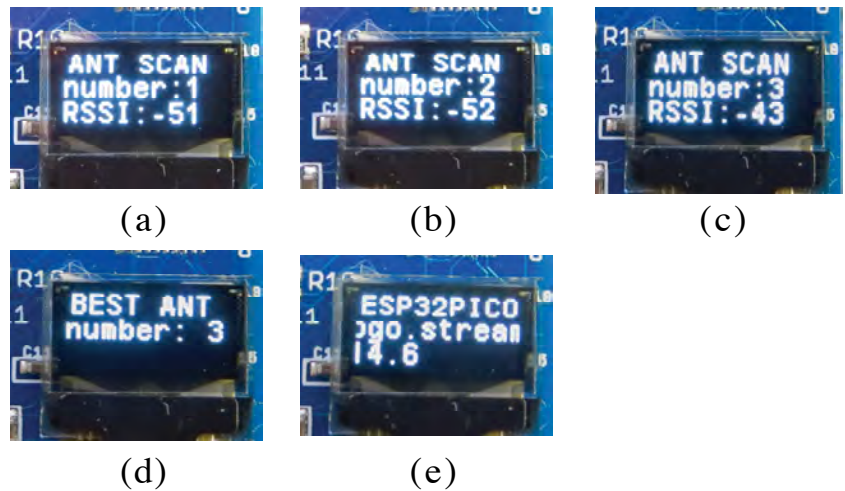

(a)　　　　　　(b)　　　　　　(c)

(d)　　　　　　(e)

Fig. 10 An example of actual antenna selecting process. Figures (a), (b) and (c) are showing RSSI of a given access point defined by SSID for antenna number "n", and (d) determines the best antenna for communication, and (e) shows the web-radio process starting.

Figures 10 show the actual example of the board operation. Fig. 10 (a) to (c) show the antenna switching operation, where RSSI are presented on OLED display by dB scale for each antenna selection case. It is not presented on the OLED display but these RSSI values are measured for the given SSID, corresponding one Wi-Fi access point link status. After switching antenna, the best antenna is determined (in this case, antenna #3) and starting web radio process. By the implementation by n24bass [10], the second line of the display shows the current receiving web radio station URL and the third line shows the IP address of the board, which is delivered by DHCP. The screen is not wide enough to show full information but the n24bass implementation realizes character scrolling to leftward.
The full demo movie, from power on to music reception is can be found at
https://youtu.be/tu75nzV8ipI